\documentclass[12pt,a4paper]{article}

\usepackage{cite}
\usepackage[utf8]{inputenc}
\usepackage[english]{babel}
\usepackage{amsmath}
\usepackage{mathrsfs}
\usepackage{amsfonts}
\usepackage{amssymb}
\usepackage{makeidx}
\usepackage{graphics}
\usepackage{float}
\usepackage{subfloat}
\usepackage{subfigure}
\usepackage{wrapfig}
\usepackage[usenames, dvipsnames]{color}
\usepackage{tcolorbox}
\usepackage{hyperref}
\definecolor{link}{rgb}{0.1,0.1,0.9}
\hypersetup{colorlinks=true,linkcolor=link,citecolor=link,urlcolor=link,linktocpage}

\usepackage[left=1.8cm,right=1.8cm,top=2cm,bottom=2cm]{geometry}
\usepackage[capitalize]{cleveref}
\usepackage{tikz}
\usetikzlibrary{shapes}
\usetikzlibrary{plotmarks}
\makeatletter
\def\Let@{\def\\{\notag\math@cr}}
\makeatother

\usepackage{authblk}
\newcommand{\be}{\begin{equation}}
\newcommand{\ee}{\end{equation}}
\newcommand{\ben}{\begin{eqnarray}\displaystyle}
\newcommand{\een}{\end{eqnarray}}

\newcommand{\bea}[1]{\begin{eqnarray}\label{#1} }
\newcommand{\eea}{\end{eqnarray}}

\newcommand{\Qp}{\mathbb{Q}_p}
\newcommand{\Zp}{\mathbb{Z}_p}

\def\boxempty{{\,\lower0.9pt\vbox{\hrule \hbox{\vrule height 0.25 cm
\hskip 0.25 cm \vrule height 0.25 cm}\hrule}\,}}

\begin{document}
%%%%%%%%%%%%%%%%%%%%%%%%%%%%%%%%%%%%%%%%%%
\begin{titlepage}

\title{
{\Huge\bf On the Exchange Interactions}\\ 
{\Huge\bf in Holographic $p$-adic CFT}
}

\bigskip\bigskip\bigskip\bigskip\bigskip

\author{{\bf Parikshit Dutta}${}^{1,2}$\thanks{{\tt parikshitdutta@yahoo.co.in}},
                    {\bf Debashis Ghoshal}${}^1$\thanks{{\tt dghoshal@mail.jnu.ac.in}}, \\  
                        {\bf and Arindam Lala}$^1$\thanks{{\tt arindam.physics1@gmail.com}}\\
\hfill\\              
${}^1${\it School of Physical Sciences, Jawaharlal Nehru University,}\\
{\it New Delhi 110067, India}\\
\hfill\\
${}^2${\it Asutosh College, 92 Shyama Prasad Mukherjee Road,}\\
{\it Kolkata 700026, India}
     }

\iffalse
\author{
{\large\bf Parikshit Dutta}\\
\\
{}\\
{\large\bf Debashis Ghoshal}\\
{\it Dipartimento di Fisica, Universit\`{a} degli studi di Genova}\\
%{\large\it 16146 Genova, Italy}\\
and\\
{\it Asutosh College} \\
{\it  92 Shyama Prasad Mukherjee Road}\\ 
{\it Kolkata 700026, India}\\
{\tt pdutta@yahoo.co.in}\\
{}\\
{\large\bf Arindam Lala}\\
{\it School of Physical Sciences, Jawaharlal Nehru University}\\
{\it New Delhi 110067, India}\\

}
\fi

\bigskip\bigskip\bigskip\bigskip

\date{%
%{\bf PACS: 11.10.Lm, 11.25.Sq, 87.10.Ed}
%%
%%
%%  11.10.Lm (Nonlinear or nonlocal theories and models)
%%  (see also 11.27.+d Extended classical solutions; cosmic strings,
%%  domain walls, texture)
%%
%%  11.25.Sq  (Nonperturbative techniques; string field theory )
%%
%%  87.10.Ed (Ordinary differential equations (ODE), partial
%%  differential equations (PDE), integrodifferential models)
%%
%%
%
\bigskip\bigskip
\begin{quote}
\centerline{{\bf Abstract}}
{\small
There is a renewed interest in conformal field theories (CFT) on ultrametric spaces 
($p$-adic field and its algebraic extensions) in view of their natural adaptability in the
holographic setting. We compute the contributions from the exchange interactions to
the four-point correlator of the CFT using Witten diagrams with three-scalar interaction 
vertex. Together with the contributions from the bulk four-point interaction, the contact 
term, these provide a complete answer. We remark on the singularity structure in Mellin 
space, and argue that all these models are analogues of adS$_2$/CFT$_1$.
}
\end{quote}
}

\bigskip
%\leftline{{\bf Report No: }} }

%\begin{abstract}
%\end{abstract}

\end{titlepage}
%%%%%%%%%%%%%%%%%%%%%%%%%%%%%%%%%%%%%%%%%%
\thispagestyle{empty}\maketitle\vfill \eject

\tableofcontents
%\newpage

%%%%%%%%%%%%%%%%%%%%%%%%%%%%%%%%%%%%%%%%%%
\section{Introduction}\label{sec:Introd}
Conformally invariant field theories (CFT) are distinguished in the space of quantum field 
theories, and are of fundamental importance in understanding their dynamics. The holographic
duality \cite{Maldacena:1997re,Witten:1998qj,Gubser:1998bc} offers a different approach to 
analysing CFTs. It is of interest to study CFTs based on different spaces to have a better handle 
on their underlying mathematical structure. This has been the motivation to explore CFTs based 
on ultrametric fields ($p$-adic fields $\Qp$, for a fixed prime $p$, and its algebraic extensions 
${\mathbb{K}}$ of $\Qp$) \cite{Melzer:1988he,Spokoiny:1988zk,Parisi:1988yc,Zhang:1988ku}). 
Recently, Gubser {\em et al} \cite{Gubser:2016guj} (see also \cite{Heydeman:2016ldy}) initiated a 
study of CFTs on ultrametric spaces as holographic dual to bulk theories on uniform lattices 
(Bruhat-Tits trees). In the simplest setting, the bulk theory is that of a scalar field
\cite{Muck:1998rr,DHoker:1999kzh}, defined by a lattice action \cite{Zabrodin:1988ep}. The 
correlators of the boundary CFT are computed by evaluating Witten diagrams in the bulk 
theory. Specifically, the two-, three- and four-point correlation functions in the ultrametric 
CFT were calculated and shown to have structures very similar to the usual (Archimedean) 
CFTs. There are some key differences, however, arising from the absence of descendent 
fields, which in turn may be attributed to the vanishing of local derivatives along the ultrametric 
boundary.

Specifically, the four-point correlation function in Ref.\cite{Gubser:2016guj} was computed with 
a four-point contact interaction in the bulk. In this paper, we compute the contribution to the 
same correlator from the exchange diagrams resulting from a three-point interaction vertex in 
the bulk\footnote{In fact, these contributions have been considered recently in 
Ref.\cite{Gubser:2017tsi}. However, the authors reduce these to the `geodesic bulk diagrams' 
with the objective to investigate the conformal blocks. We have done a direct computation. 
The sum of various configurations evaluated in the following should contribute to a conformal 
block.}. 

We also argue that all such ultrametric CFTs (on $\Qp$ or its algebraic extensions) are really 
like {\em one-dimensional} CFTs, that provide discrete analogues of the holographic duality 
adS$_2$/CFT$_1$. Finally, we analyze the Mellin space representation of the four-point 
function.

The projective $p$-adic line $\Qp\mathbb{P}^1$ is the asymptotic boundary of an infinite 
lattice, the Bruhat-Tits tree $\mathcal{T}_p$ (more familiar to physicists as Bethe lattice with 
coordination number $p+1$) that has been used as the `worldsheet' of open $p$-adic string 
\cite{Zabrodin:1988ep,Chekhov:1989bg}. This is exactly analogous to the real case. The 
group action GL($2,\Qp$) naturally leads to the symmetric space $\mathcal {T}_p$ obtained 
by a quotient of its maximal compact subgroup GL($2,\Zp$), in which the entries are $p$-adic 
integers, Ultrametric spaces that are finite algebraic extensions of $\Qp$ have a similar 
structure (possibly with different coordination number). Moreover, one may think of a family 
of so called ramified extensions of $\Qp$ as a continuum limit\cite{Ghoshal:2006te}.

%%%%%%%%%%%%%%%%%%%%%%%%%%%%%%%%%%%%%%%%%%
\section{A brief review of ultrametric holography}\label{sec:UMHol}
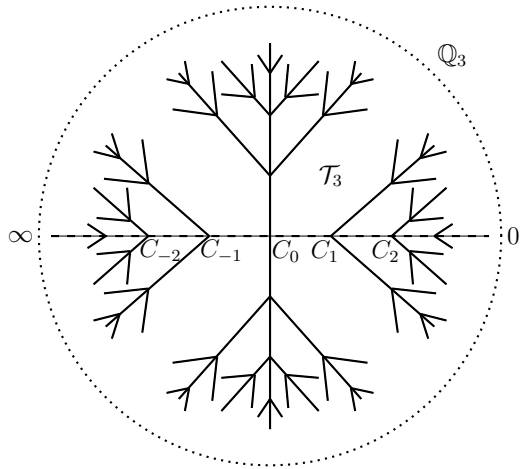
\begin{figure}[H]
\centering
\begin{tikzpicture}[scale=0.8,every node/.style={scale=0.8}]
%\draw[step=1cm,gray,thin] (-3,-3) grid (3,3);
\draw [black,dotted,thick](0,0) circle (3.8cm) ;
\filldraw[thick] (0,3.2) -- (0,-3.2);
\filldraw[thick,dashed] (-3.6,0) -- (3.6,0);
\filldraw[thick] (1,0) -- (2.7,-1.5);
\filldraw[thick] (1,0) -- (2.7,1.5);
%%%%%%%%%%%%%%%%%%%%%%%%%%%%%
\filldraw[thick] (2,0) -- (2.9,.8);
\filldraw[thick] (2,0) -- (2.9,-.8);

\filldraw[thick] (2.7,0) -- (3,0.2);
\filldraw[thick] (2.7,0) -- (3,-0.2);

\filldraw[thick] (2,0.87) -- (2.1,1.6);
\filldraw[thick] (2,0.87) -- (2.73,0.95);

\filldraw[thick] (2.47,1.29) -- (2.6,1.7);
\filldraw[thick] (2.47,1.29) -- (2.9,1.4);

\filldraw[thick] (2.47,-1.29) -- (2.6,-1.7);
\filldraw[thick] (2.47,-1.29) -- (2.9,-1.4);

\filldraw[thick] (2.3,0.25) -- (2.35,0.8);
\filldraw[thick] (2.3,0.25) -- (2.85,0.3);

\filldraw[thick] (2,-.87) -- (2.1,-1.6);
\filldraw[thick] (2,-0.87) -- (2.73,-0.95);

\filldraw[thick] (2.3,-0.25) -- (2.35,-0.8);
\filldraw[thick] (2.3,-0.25) -- (2.85,-0.3);
%%%%%%%%%%%%%%%%%%%%%%%%%%%%%%%%%%%%%%
\filldraw[thick] (-1,0) -- (-2.7,-1.5);
\filldraw[thick] (-1,0) -- (-2.7,1.5);

\filldraw[thick] (-2,0) -- (-2.9,.8);
\filldraw[thick] (-2,0) -- (-2.9,-.8);

\filldraw[thick] (-2.7,0) -- (-3,0.2);
\filldraw[thick] (-2.7,0) -- (-3,-0.2);

\filldraw[thick] (-2,0.87) -- (-2.1,1.6);
\filldraw[thick] (-2,0.87) -- (-2.73,0.95);

\filldraw[thick] (-2.47,1.29) -- (-2.6,1.7);
\filldraw[thick] (-2.47,1.29) -- (-2.9,1.4);

\filldraw[thick] (-2.47,-1.29) -- (-2.6,-1.7);
\filldraw[thick] (-2.47,-1.29) -- (-2.9,-1.4);

\filldraw[thick] (-2.3,0.25) -- (-2.35,0.8);
\filldraw[thick] (-2.3,0.25) -- (-2.85,0.3);

\filldraw[thick] (-2,-.87) -- (-2.1,-1.6);
\filldraw[thick] (-2,-0.87) -- (-2.73,-0.95);

\filldraw[thick] (-2.3,-0.25) -- (-2.35,-0.8);
\filldraw[thick] (-2.3,-0.25) -- (-2.85,-0.3);
%%%%%%%%%%%%%%%%%%%%%%%%%%%%%%%%%%%%%%%%%%
\filldraw[thick] (0,1) -- (1.5,2.7);
\filldraw[thick] (0,1) -- (-1.5,2.7);

\filldraw[thick] (0,2) -- (.8,2.9);
\filldraw[thick] (0,2) -- (-.8,2.9);

\filldraw[thick] (0,2.7) -- (0.2,3);
\filldraw[thick] (0,2.7) -- (-0.2,3);

\filldraw[thick] (0.87,2) -- (1.6,2.1);
\filldraw[thick] (0.87,2) -- (0.95,2.73);

\filldraw[thick] (1.33,2.52) -- (1.7,2.6);
\filldraw[thick] (1.33,2.52) -- (1.4,2.9);

\filldraw[thick] (-1.33,2.52) -- (-1.7,2.6);
\filldraw[thick] (-1.33,2.52) -- (-1.4,2.9);

\filldraw[thick] (-0.25,2.3) -- (-0.8,2.35);
\filldraw[thick] (-0.25,2.3) -- (-0.3,2.85);

\filldraw[thick] (-.87,2) -- (-1.6,2.1);
\filldraw[thick] (-0.87,2) -- (-0.95,2.73);

\filldraw[thick] (0.25,2.3) -- (0.8,2.35);
\filldraw[thick] (0.25,2.3) -- (0.3,2.85);
%%%%%%%%%%%%%%%%%%%%%%%%%%%%%%%%%%%%%%%%%%
\filldraw[thick] (0,-1) -- (1.5,-2.7);
\filldraw[thick] (0,-1) -- (-1.5,-2.7);

\filldraw[thick] (0,-2) -- (.8,-2.9);
\filldraw[thick] (0,-2) -- (-.8,-2.9);

\filldraw[thick] (0,-2.7) -- (0.2,-3);
\filldraw[thick] (0,-2.7) -- (-0.2,-3);

\filldraw[thick] (0.87,-2) -- (1.6,-2.1);
\filldraw[thick] (0.87,-2) -- (0.95,-2.73);

\filldraw[thick] (1.33,-2.52) -- (1.7,-2.6);
\filldraw[thick] (1.33,-2.52) -- (1.4,-2.9);

\filldraw[thick] (-1.33,-2.52) -- (-1.7,-2.6);
\filldraw[thick] (-1.33,-2.52) -- (-1.4,-2.9);

\filldraw[thick] (-0.25,-2.3) -- (-0.8,-2.35);
\filldraw[thick] (-0.25,-2.3) -- (-0.3,-2.85);

\filldraw[thick] (-.87,-2) -- (-1.6,-2.1);
\filldraw[thick] (-0.87,-2) -- (-0.95,-2.73);

\filldraw[thick] (0.25,-2.3) -- (0.8,-2.35);
\filldraw[thick] (0.25,-2.3) -- (0.3,-2.85);

\node at (0.26,-0.25) {$C_0$};
\node at (.9,-0.25) {$C_{1}$};
\node at (1.9,-0.25) {$C_{2}$};
\node at (4.0,0) {$0$};

\node at (-.8,-0.25) {$C_{-1}$};
\node at (-1.8,-0.25) {$C_{-2}$};
\node at (-4.1,0) {$\infty$};
\node at (1.0,1.0) {$\mathcal{T}_3$};
\node at (3.0,3.0) {$\mathbb{Q}_3$};
\end{tikzpicture}
\caption{{\small
A part of the Bruhat-Tits tree corresponding to the 3-adic field $\mathbb{Q}_3$ at its asymptotic
boundary. The dotted line connects 0 to $\infty$. The label on the node $C_N$ corresponds to 
the leading power in the Laurent expansion of $x$, and the branching is decided by the 
coefficients $a_n$.
}}
\label{fig:tree3}
\end{figure}

The set up we consider is that of Ref.\cite{Gubser:2016guj}, therefore, we shall be 
brief. A $p$-adic number $x\in\Qp$ (for a fixed prime $p$) has a Laurent series 
expansion\footnote{We shall not recount any aspect of ultrametric analysis since succinct 
accounts of the facts relevant to us exist in many physics articles --- see, e.g., Appendix 
A of \cite{Ghoshal:2004ay} or Sec.~2 of \cite{Gubser:2016guj}, or \cite{Brekke:1993gf} for 
a more comprehensive account.} in $p$ 
\begin{equation*}
x = p^{N}\left(a_0 + a_1 p + a_2 p^2 + a_3 p^3 + \cdots\right),\qquad a_0\ne 0,\: 
a_n=\left\{0,1,2,\cdots,p-1\right\}\:\mbox{\rm and } N\in\mathbb{Z}.
\end{equation*}
Its ($p$-adic) norm is $\left|x\right|_p = p^{-N}$. The above may be used to construct the 
bulk, a Bethe lattice with coordination number $p+1$ as shown in \cref{fig:tree3}. It is also
the equivalence classes of lattices in $\Qp^2$. An algebraic extension $\mathbb{K}$ of it 
is a finite dimensional vector space over $\Qp$. It is also an ultrametric space that share 
all qualitative features of $\Qp$. In particular, there is an associated Bruhat-Tits tree (which 
may be constructed as a symmetric space from the group action on $\mathbb{K}$, with its 
coordination number depending on $p$ and the ramification index of the extension). What 
is important, however, is the fact that it is still a tree, an infinite graph with no closed loop. 
Therefore, the path between any two points, either in the bulk or on its asymptotic boundary, 
is unique (discounting backtracking). 

As a result, contrary to what one may expect from the dimension $d$ of (an unramified 
extension) $\mathbb{K}$ as a $\Qp$-vector space, which may lead one believe that the 
boundary CFT on $\mathbb{K}$ is $d$-dimensional, it is still only an analogue of a {\em 
one-dimensional CFT}. This fact is consistent with what we shall find for the boundary 
$n$-point functions: they depend on $(n-3)$ cross ratios, as in one-dimensional 
Archimedean CFTs \cite{Penedones:2016voo}. There is further support from the fact 
that GL(2,$\mathbb{K}$), the group of symmetries, of the CFT is rather akin to 
SL(2,$\mathbb{R}$). After some confusion in the early days of ultrametric string 
theory, those based on algebraic extensions of $\Qp$ were also realised to be 
analogues of open string theory for which the boundary of the worldsheet is 
one-dimensional \cite{Zabrodin:1988ep}.

It would suffice, therefore, to restrict to the simplest case on $\Qp$. Nevertheless to 
facilitate comparison with the existing literature, we shall present the contributions 
from the bulk three-vertex to the boundary CFT correlators for the general case of an 
$n$-dimensional (unramified) extension $\mathbb{K}$ of $\Qp$. Essentially, this amounts 
to replacing $p$ by $q\equiv p^n$ in some of the formulas \cite{Gubser:2016guj}. To 
get back to $p$-adic CFT one sets $n=1$.

%%%%%%%%%%%%%%%%%%%%%%%%%%%%%%%%%%
%%%%%%%%%%%%%%%%%%%%%%%%%%%%%%%%%%

The bulk theory is a scalar field theory defined by the discrete lattice action
\begin{equation*}
S[\phi] = \frac{1}{2} \sum_{E_{\langle ab\rangle}\in \mathcal{T}_p}
\left(\phi_a - \phi_b\right)^2 + \sum_{a\in\mathcal{T}_p} \left( \frac{1}{2} m^2\phi_a^2
+ \frac{1}{3!} g_3\phi_a^3 + \frac{1}{4!} g_4 \phi_a^4\right)
\end{equation*}
where the first sum is over all edges $E_{\langle ab\rangle}$ and the second over all nodes
of the tree. Using the fact that the tree is the equivalence classes of two-dimensional lattices, 
Ref.\cite{Gubser:2016guj} labelled a bulk point $a\in\mathcal{T}_p$ by a pair of $p$-adic 
numbers $a=(r,x)$, where $x\in\Qp$ and $r = p^m\in\Qp$, ($m\in\mathbb{Z}$). This denotes 
an equivalence class of the form $x + r\mathbb{Z}_p$. One may think of $r$ as the `radial' 
coordinate related to the notion of a `depth' from the `trunk' joining 0 to $\infty$. 

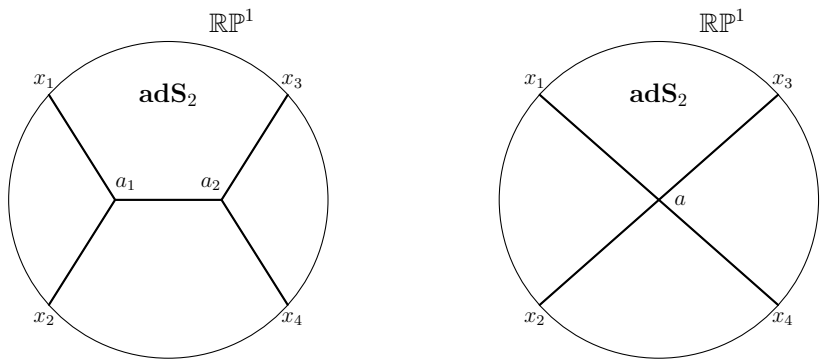
\begin{figure}[H]
\centering
\subfigure{
\begin{tikzpicture}[scale=.7,every node/.style={scale=0.7}]
%\draw[step=1cm,gray,thin] (-3,-3) grid (3,3);
\draw (0,0) circle (3cm) ;
\filldraw[thick] (-1,0) -- (1,0);
\filldraw[thick] (-1,0) -- (-2.25,2);
\filldraw[thick] (-1,0) -- (-2.25,-2);
\filldraw[thick] (1,0) -- (2.25,2);
\filldraw[thick] (1,0) -- (2.25,-2);
%\filldraw[thick] (-1.2,-.3) -- (-2,.5);
%\filldraw[thick] (-1.6,-.95) -- (-2.55,-.1);
\node[mark size=2pt,color=red] at (-1,0) {\pgfuseplotmark{*}};
%\node[mark size=2pt,color=blue] at (-1.2,-.3) {\pgfuseplotmark{*}};
%\node[mark size=2pt,color=blue] at (-1.6,-.95) {\pgfuseplotmark{*}};
%\node[mark size=2pt,color=blue] at (-2,.5) {\pgfuseplotmark{*}};
%\node[mark size=2pt,color=blue] at (-2.55,-.1) {\pgfuseplotmark{*}};
\node[mark size=2pt,color=red] at (1,0) {\pgfuseplotmark{*}};
\node at (-.8,0.3) {$a_{1}$};
\node at (.8,0.3) {$a_{2}$};
\node at (0,2.0) {\large \textbf{adS$_2$}};
\node at (1.2,3.4) {\large $\mathbb{RP}^{1}$};
%\node at (-2.3,.55) {$a_{2}$};
%\node at (-2.6,.2) {$a_{1}$};
\node at (-2.33,2.25) {$x_{1} $};
\node at (-2.33,-2.25) {$x_{2} $};
\node at (2.33,2.25) {$x_{3} $};
\node at (2.33,-2.25) {$x_{4} $};
\end{tikzpicture}
}
\hspace{48pt}
\subfigure{
\begin{tikzpicture}[scale=.7,every node/.style={scale=0.7}] 
%\draw[step=1cm,gray,thin] (-3,-3) grid (3,3);
\draw (0,0) circle (3cm) ;
%\filldraw[thick] (-1,0) -- (1,0);
\filldraw[thick] (0,0) -- (-2.25,2);
\filldraw[thick] (0,0) -- (-2.25,-2);
\filldraw[thick] (0,0) -- (2.25,2);
\filldraw[thick] (0,0) -- (2.25,-2);
%\filldraw[thick] (-.4,0) -- (-.4,1.3);
%\filldraw[thick] (.4,0) -- (.4,1.3);
\node[mark size=2pt,color=red] at (0,0) {\pgfuseplotmark{*}};
%\node[mark size=2pt,color=blue] at (-.4,0) {\pgfuseplotmark{*}};
%\node[mark size=2pt,color=blue] at (-.4,1.3) {\pgfuseplotmark{*}};
%\node[mark size=2pt,color=blue] at (.4,0) {\pgfuseplotmark{*}};
%\node[mark size=2pt,color=blue] at (.4,1.3) {\pgfuseplotmark{*}};
%\node[mark size=2pt,color=red] at (1,0) {\pgfuseplotmark{*}};
\node at (0.4,0) {$a$};
%\node at (.8,0.3) {$c_{2}$};
\node at (0,2.0) {\large \textbf{adS$_2$}};
\node at (1.2,3.4) {\large $\mathbb{RP}^{1}$};
%\node at (.4,1.6) {$a_{2}$};
%\node at (-.4,1.6) {$a_{1}$};
\node at (-2.33,2.25) {$x_{1} $};
\node at (-2.33,-2.25) {$x_{2} $};
\node at (2.33,2.25) {$x_{3} $};
\node at (2.33,-2.25) {$x_{4} $};
\end{tikzpicture}
}
\caption{\small Representative Witten diagrams contributing to the four-point correlators in 
Archimedean adS$_{2}$/CFT$_{1}$ from bulk three- and four-point interaction vertices.
The correlators involve integrating over the bulk points $a_1$, $a_2$ and $a$.}
\label{fig:ArchWittenDiag}
\end{figure}

In the following, we shall compute the contribution of the exchange diagrams resulting from
the $\phi^3$ interaction to the boundary four-point correlation function. Together with the 
contact term contribution from the $\phi^4$ interaction, computed in \cite{Gubser:2016guj}, 
this accounts for all the contributions to the bulk four-point amplitude. The corresponding 
Witten diagrams in the usual Archimedean adS$_2$/CFT$_1$ are shown in 
\cref{fig:ArchWittenDiag}.

%%%%%%%%%%%%%%%%%%%%%%%%%%%%%
%%%%%%%%%%%%%%%%%%%%%%%%%%%%%%%%%%%%%%%%%%%%

In order to evaluate the contribution of a Witten diagram W, one needs the propagators. The 
(unnormalized) bulk-to-bulk propagator $G(a,b)$ ($a,b\in\mathcal{T}_p$) is the Green's function 
of the free massive field
\begin{equation}
G(a_1,a_2) 
%= \zeta_p(2\Delta) p^{-\left(d(a_1,a_2)+1\right)\Delta},   \label{BBProp}
=  p^{-d(a_1,a_2)\Delta},   \label{BBProp}
\end{equation}
where $d(a_1,a_2)$ is the number of edges connecting the nodes and $\Delta$ is related to 
the mass by the local zeta function at the prime $p$, $\zeta_p(s) =1/\left(1-p^{-s}\right)$ through 
$m^{-2} = -\,\zeta_p(\Delta - n)\zeta_p(-\Delta)$. The bulk-to-boundary propagator $K(a,x)$ was 
obtained as a regularised limit of sending one of the points in the(unnormalized)  bulk-to-bulk 
propagator to the boundary, and is expressed as
\begin{equation}
K(a,x) \equiv K\!\left((r_y,y) , x\right)) = 
%\frac{\zeta_{p}(2\Delta)}{\zeta_{p}(2\Delta-1)}\,
\frac{\left|r_y\right|_p^\Delta}{\left(\sup\left\{\left|r_y\right|_p,\left|y-x)\right|_q\right\}\right)^{2\Delta}}.
\label{B2bProp}
\end{equation}
These propagators are similar to those in Archimedean adS/CFT. 

Let us consider the four-point correlation function of the boundary operator $\mathcal{O}$ of
dimension $\Delta$, dual to the field $\phi$. Using the GL(2,$\mathbb{K}$) symmetry, the 
positions of three insertions in $\mathcal{A}_4(x_1,x_2,x_3,x_4) = \big\langle \mathcal{O}(x_1)
\mathcal{O}(x_2)\mathcal{O}(x_3)\mathcal{O}(x_4)\big\rangle$ can be sent to three 
predetermined points. Hence, modulo dependence fixed by scaling, it is a function of the 
cross ratio
\begin{equation}
u = \left| \frac{(x_1-x_2)(x_3-x_4)}{(x_1-x_3)(x_2-x_4)}\right|_q = p^{-d(c_1,c_2)}.
\label{xratio}
\end{equation}
The second expression for the cross ratio involves the distance between the bulk points $c_1$ 
and $c_2$, which are the two ends of the segment along which the path from $x_1$ to $x_3$ 
and $x_2$ to $x_4$ overlap. It is important that these paths are {\em unique}, and all paths 
(without any backtracking) on the tree are geodesics. The four point function in a 
$d$-dimensional CFT depends on {\em two} independent cross ratios $u$ and $v$ for 
$d\ge 2$ \cite{Penedones:2016voo}. The $p$-adic adS/CFT, however, is analogous to the 
case $d=1$, where there is only {\em one} independent cross ratio. It is clear that once the 
three insertions are fixed, say at $x_1=0$, $x_3=1$ and $x_4\to\infty$, inserting the fourth 
one anywhere leads to only one independent cross ratio.  

\begin{figure}[h]
\centering
\subfigure{
\begin{tikzpicture}[scale=0.7,every node/.style={scale=0.7}]
%\draw[step=1cm,gray,thin] (-3,-3) grid (3,3);
\draw (0,0) circle (3cm) ;
\filldraw[thick] (-1,0) -- (1,0);
\filldraw[thick] (-1,0) -- (-2.25,2);
\filldraw[thick] (-1,0) -- (-2.25,-2);
\filldraw[thick] (1,0) -- (2.25,2);
\filldraw[thick] (1,0) -- (2.25,-2);
%\filldraw[thick] (-1.6,-.95) -- (-.6,-1.5);
\filldraw[thick] (0,0) -- (0,1.2);
%\filldraw[thick] (-.6,-1.5) -- (0,-2.4);
\node[mark size=2pt,color=red] at (-1,0) {\pgfuseplotmark{*}};
\node[mark size=2pt,color=blue] at (0,0) {\pgfuseplotmark{*}};
%\node[mark size=2pt,color=blue] at (-1.6,-.95) {\pgfuseplotmark{*}};
\node[mark size=2pt,color=blue] at (0,1.2) {\pgfuseplotmark{*}};
%\node[mark size=2pt,color=blue] at (-.6,-1.5) {\pgfuseplotmark{*}};
\node[mark size=2pt,color=red] at (1,0) {\pgfuseplotmark{*}};
\node at (-.8,0.3) {$c_{1}$};
\node at (.8,0.3) {$c_{2}$};
%\node at (-.6,-1.2) {$a_{1}$};
%\node at (-1.7,-.6) {$b_{1}$};
\node at (.3,1) {$a$};
\node at (0,-.35) {$b$};
\node at (0,-1.75) {$\mathcal{T}_p$};
\node at (3.85,0.0) {$\Qp\mathbb{P}^1$};
\node at (-2.33,2.25) {$x_{1} $};
\node at (-2.33,-2.25) {$x_{2} $};
\node at (2.33,2.25) {$x_{3} $};
\node at (2.33,-2.25) {$x_{4} $};
\end{tikzpicture}
}
\hspace{48pt}
\subfigure{
\begin{tikzpicture}[scale=0.7,every node/.style={scale=0.7}]
%\draw[step=1cm,gray,thin] (-3,-3) grid (3,3);
\draw (0,0) circle (3cm) ;
\filldraw[thick] (-1,0) -- (1,0);
\filldraw[thick] (-1,0) -- (-2.25,2);
\filldraw[thick] (-1,0) -- (-2.25,-2);
\filldraw[thick] (1,0) -- (2.25,2);
\filldraw[thick] (1,0) -- (2.25,-2);
\filldraw[thick] (-1.6,-.95) -- (-.6,-1.5);
%\filldraw[thick] (0,0) -- (0,1.2);
%\filldraw[thick] (-.6,-1.5) -- (0,-2.4);
\node[mark size=2pt,color=red] at (-1,0) {\pgfuseplotmark{*}};
%\node[mark size=2pt,color=blue] at (0,0) {\pgfuseplotmark{*}};
\node[mark size=2pt,color=blue] at (-1.6,-.95) {\pgfuseplotmark{*}};
%\node[mark size=2pt,color=blue] at (0,1.2) {\pgfuseplotmark{*}};
\node[mark size=2pt,color=blue] at (-.6,-1.5) {\pgfuseplotmark{*}};
\node[mark size=2pt,color=red] at (1,0) {\pgfuseplotmark{*}};
\node at (-.8,0.3) {$c_{1}$};
\node at (.8,0.3) {$c_{2}$};
\node at (-.6,-1.2) {$a$};
\node at (-1.7,-.6) {$b$};
\node at (0,1.75) {$\mathcal{T}_p$};
\node at (3.85,0.0) {$\Qp\mathbb{P}^1$};
%\node at (.3,1) {$a$};
%\node at (0,-.3) {$b$};
\node at (-2.33,2.25) {$x_{1} $};
\node at (-2.33,-2.25) {$x_{2} $};
\node at (2.33,2.25) {$x_{3} $};
\node at (2.33,-2.25) {$x_{4} $};
\end{tikzpicture}
}
\caption{\small Examples of Witten (subway) diagrams with bulk 4-point contact
interaction that contribute to the CFT  four-point correlator. The position of the 
interaction point $a$ is integrated over the tree $\mathcal{T}_p$.}
\label{fig:Gubser4Subway}
\end{figure}
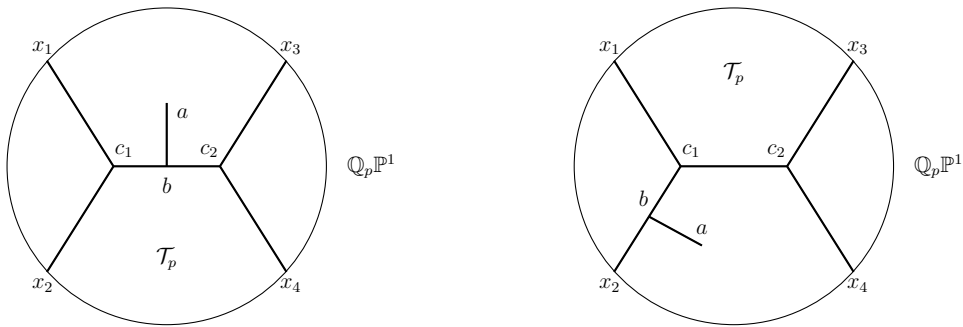

The contribution to $\mathcal{A}_4^{(4)}$ from the contact interaction, computed in 
\cite{Gubser:2016guj}, is 
\begin{equation}
\mathcal{A}_4^{(4)} = g_{4} \, u^{2\Delta}\,
\left[\frac{\log_{p} u}{\zeta_{p}(4\Delta)} - \left(\frac{\zeta_{p}(2\Delta)}{\zeta_{p}(4\Delta)}+1\right)^{2} + 3\right]
\times\frac{\zeta_{p}(4\Delta -n)}{\left(\left|x_{12}\right|_q \left|x_{34}\right|_q\right)^{2\Delta}}  
\label{contact4} 
\end{equation}
Some of the Witten diagrams on the tree, dubbed {\em subway diagrams} in \cite{Gubser:2016guj}, are 
shown in \cref{fig:Gubser4Subway}. In spite of their deceptive appearance, these are contact diagrams. 

%%%%%%%%%%%%%%

%%%%%%%%%%%%%%%%%%%%%%%%%%%%%%%%%%%%%%%%
\section{Boundary correlators with bulk 3-point interaction}\label{sec:A4_3vertex}
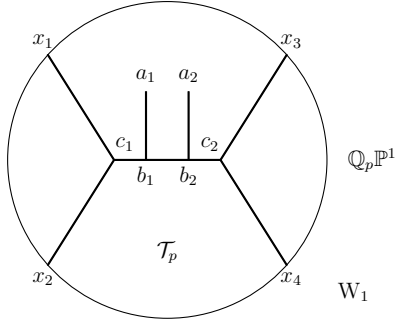
\begin{figure}[h]
\centering
\subfigure{
\begin{tikzpicture}[scale=0.7,every node/.style={scale=0.7}]
%\draw[step=1cm,gray,thin] (-3,-3) grid (3,3);
\draw (0,0) circle (3cm) ;
\filldraw[thick] (-1,0) -- (1,0);
\filldraw[thick] (-1,0) -- (-2.25,2);
\filldraw[thick] (-1,0) -- (-2.25,-2);
\filldraw[thick] (1,0) -- (2.25,2);
\filldraw[thick] (1,0) -- (2.25,-2);
\filldraw[thick] (-.4,0) -- (-.4,1.3);
\filldraw[thick] (.4,0) -- (.4,1.3);
\node[mark size=2pt,color=red] at (-1,0) {\pgfuseplotmark{*}};
\node[mark size=2pt,color=blue] at (-.4,0) {\pgfuseplotmark{*}};
\node[mark size=2pt,color=blue] at (-.4,1.3) {\pgfuseplotmark{*}};
\node[mark size=2pt,color=blue] at (.4,0) {\pgfuseplotmark{*}};
\node[mark size=2pt,color=blue] at (.4,1.3) {\pgfuseplotmark{*}};
\node[mark size=2pt,color=red] at (1,0) {\pgfuseplotmark{*}};
\node at (-.8,0.3) {$c_{1}$};
\node at (.8,0.3) {$c_{2}$};
\node at (.4,-0.3) {$b_{2}$};
\node at (-.4,-0.3) {$b_{1}$};
\node at (.4,1.6) {$a_{2}$};
\node at (-.4,1.6) {$a_{1}$};
\node at (-2.33,2.25) {$x_{1} $};
\node at (-2.33,-2.25) {$x_{2} $};
\node at (2.33,2.25) {$x_{3} $};
\node at (0,-1.75) {$\mathcal{T}_p$};
\node at (3.85,0.0) {$\Qp\mathbb{P}^1$};
\node at (2.33,-2.25) {$x_{4} $};
\node at (3.5,-2.5) {W$_1$};
\end{tikzpicture}
}
\caption{\small
Examples of Witten (subway) diagrams for the four-point CFT correlator with bulk 3-point 
exchange interactions. In this configuration both the bulk points project on the common 
segment joining the boundary points, with $b_1$ always to the left of $b_2$.}
\label{subway1-4pf}
\end{figure}

\begin{figure}[h]
\centering
\subfigure{
\begin{tikzpicture}[scale=0.7,every node/.style={scale=0.7}]
%\draw[step=1cm,gray,thin] (-3,-3) grid (3,3);
\draw (0,0) circle (3cm) ;
\filldraw[thick] (-1,0) -- (1,0);
\filldraw[thick] (-1,0) -- (-2.25,2);
\filldraw[thick] (-1,0) -- (-2.25,-2);
\filldraw[thick] (1,0) -- (2.25,2);
\filldraw[thick] (1,0) -- (2.25,-2);
\filldraw[thick] (0,0) -- (0,1);
\filldraw[thick] (0,1) -- (.6,1.9);
\filldraw[thick] (0,1) -- (-.6,1.9);
\node[mark size=2pt,color=red] at (-1,0) {\pgfuseplotmark{*}};
\node[mark size=2pt,color=blue] at (0,0) {\pgfuseplotmark{*}};
\node[mark size=2pt,color=blue] at (0,1) {\pgfuseplotmark{*}};
\node[mark size=2pt,color=blue] at (.6,1.9) {\pgfuseplotmark{*}};
\node[mark size=2pt,color=blue] at (-.6,1.9) {\pgfuseplotmark{*}};
\node[mark size=2pt,color=red] at (1,0) {\pgfuseplotmark{*}};
\node at (-2.33,2.25) {$x_{1} $};
\node at (-2.33,-2.25) {$x_{2} $};
\node at (2.33,2.25) {$x_{3} $};
\node at (2.33,-2.25) {$x_{4} $};
\node at (3.0,-2.5) {W$_2$};
\end{tikzpicture}
}
\hspace{36pt}
\subfigure{
\begin{tikzpicture}[scale=0.7,every node/.style={scale=0.7}]
%\draw[step=1cm,gray,thin] (-3,-3) grid (3,3);
\draw (0,0) circle (3cm) ;
\filldraw[thick] (-1,0) -- (1,0);
\filldraw[thick] (-1,0) -- (-2.25,2);
\filldraw[thick] (-1,0) -- (-2.25,-2);
\filldraw[thick] (1,0) -- (2.25,2);
\filldraw[thick] (1,0) -- (2.25,-2);
\filldraw[thick] (-1.6,.95) -- (-.6,1.5);
\filldraw[thick] (0,0) -- (0,1.2);
%\filldraw[thick] (-.6,-1.5) -- (0,-2.4);
\node[mark size=2pt,color=red] at (-1,0) {\pgfuseplotmark{*}};
\node[mark size=2pt,color=blue] at (0,0) {\pgfuseplotmark{*}};
\node[mark size=2pt,color=blue] at (-1.6,.95) {\pgfuseplotmark{*}};
\node[mark size=2pt,color=blue] at (0,1.2) {\pgfuseplotmark{*}};
\node[mark size=2pt,color=blue] at (-.6,1.5) {\pgfuseplotmark{*}};
\node[mark size=2pt,color=red] at (1,0) {\pgfuseplotmark{*}};
\node at (-2.33,2.25) {$x_{1} $};
\node at (-2.33,-2.25) {$x_{2} $};
\node at (2.33,2.25) {$x_{3} $};
\node at (2.33,-2.25) {$x_{4} $};
\node at (3.0,-2.5) {W$_3$};
\end{tikzpicture}
}
\caption{\small
Other Witten (subway) diagrams with bulk 3-point interactions that give
non-trivial dependence on the cross ratio.}
\label{subway2-4pf}
\end{figure}
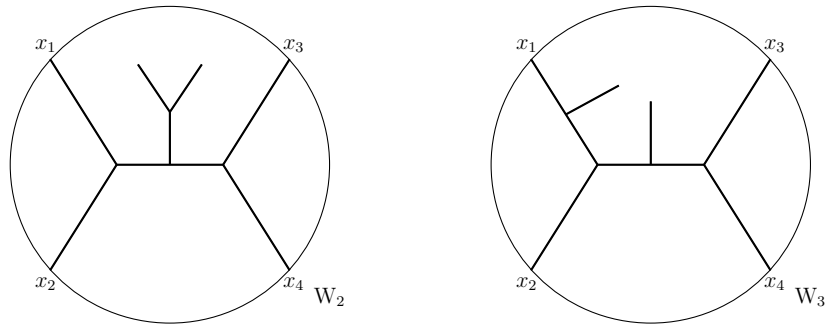

We shall now evaluate the contribution to the boundary correlators from the 3-point 
interaction vertex in the bulk. (The contact interaction will not be of importance 
in what follows, therefore, henceforth we may as well set $g_4=0$.) Let the bulk
vertices be at the nodes $a_1$ and $a_2$ of the tree $\mathcal{T}_p$, where  
$a_1$ is connected to $x_1$ and $x_2$, and $a_2$ to $x_3$ and $x_4$, by 
bulk-to-boundary propagators \cref{B2bProp}. In addition, $a_1$ and $a_2$ are 
connected by a bulk-to-bulk propagator \cref{BBProp}. Consider the (unique) paths
in $\mathcal{T}_p$ that join the boundary points. The interaction points $a_1, a_2$ 
can be joined to these paths at the points $b_1$ and $b_2$ respectively. The
path from $x_1$ to $x_3$ overlap that from $x_2$ to $x_4$ along the segment
$\langle c_1,c_2\rangle$ in the bulk. Depending on the locations of $a_1$ and 
$a_2$, which are to be summed over all the nodes, we get different configurations, 
representatives of which are shown in \cref{subway1-4pf,subway2-4pf}. The 
bulk-to-boundary propagators can now be decomposed in terms of products of 
bulk-to-boundary and bulk-to-bulk propagators in order to evaluate and simplify 
the amplitudes.

%%%%%%%%%%%%%%%%%%%%%%%%%%%%%%%%
 
 As an example, let us consider in detail the Witten diagram W$_1$ in \cref{subway1-4pf}, as 
this gives interesting new dependence on the cross ratio. Its contribution is
\begin{equation}
\mathcal{A}_4^{(3)}(\mathrm{W_1}) = g_3^2 \sum_{a_1,a_{2}\in\mathcal{T}_{p}} 
K(x_{1},a_1)  K(x_{2},a_1) K(x_{3},a_2) K(x_{4},a_2) G(a_1,a_2)
\label{4pfamp}
\end{equation}
We can decompose the bulk-to-boundary propagators as $K(x_{1},a_1) = K(x_{1},c_1)
G(c_1,b_1)G(b_1,a_1)$, and  similarly for the others. Likewise, the bulk-to-bulk propagator 
between the interaction vertices can be split as $G(a_1,a_2) = G(a_1,b_1)G(b_1,b_2)
G(b_2,a_2)$.  Notice that in this configuration, $b_1$ always to the left of $b_2$. After the 
decompositions, the contribution of this diagram can be expressed as 
\begin{equation}
\mathcal{A}_4^{(3)}(\mathrm{W_1}) = g_3^2 \left[ K(x_1,c_1) K(x_2,c_1) K(x_3,c_2)
K(x_4,c_2)\right] \sum_{a_1,a_{2}\in\mathcal{T}_{p}} 
\hat{\mathcal{A}}_4^{(3)}(\mathrm{W_1}),  \label{A43pGeneric}
\end{equation}
where we have factored out the bulk-to-boundary propagators from the boundary points to
the extreme points $c_1$ and $c_2$ of the common segment of the geodesics that join them. 

We now have to sum over the bulk points $a_1$ and $a_2$, in which sum over the different 
possibilities for $b_1$ and $b_2$ is implicit\footnote{One should take care to avoid double 
counting that may arise from degenerate cases of different configurations.}. The reduced
amplitude is 
\begin{equation*}
\hat{\mathcal{A}}_4^{(3)}(\mathrm{W}_1) = \left(G(a_1,b_1) G(a_2,b_2)\right)^3 
\left(G(c_1,b_1)G(c_2,b_2)\right)^2 G(b_1,b_2). 
\end{equation*}
Let us denote the lengths of the different segments 
as follows $l_{1}=d(a_{1},b_{1})$, $l_{2}=d(b_{2},a_{2})$, $m_1=d(c_{1},b_{1})$ and 
$m_2=d(c_2,b_2)$. The following possibilities lead to different contributions in 
$\hat{\mathcal{A}}_4^{(3)}(\mathrm{W}_1)$ as enumerated below. (In the following the 
first case occurs only if the coordination number of the lattice is at least 4.)
\begin{itemize}
\item 
$m_1=0$ and $m_2=d(c_1,c_2)$ or vice versa. This means that $b_1$ coincides with $b_2$, 
which in turn coincides with either $c_1$ or $c_2$, but the vertices $a_1$ and $a_2$ lie on 
different branches emanating from $b_1=b_2=b$. 
%%
%\begin{equation*}
\begin{align*}
& 2 u^{2\Delta}\,\left[ 2\left(p^n-2\right) \left(\sum_{l=1}^\infty p^{n(l-1)} p^{-3\Delta l}\right)
+ 2\left({p^n-2\atop 2}\right) \left(\sum_{l=1}^\infty p^{n(l-1)} p^{-3\Delta l}\right)^2\right]\\
&=\, 2u^{2\Delta} (p^n-2)p^{-3\Delta}\zeta_p(3\Delta - n)\,\left[2 + (p^n-3)
p^{-3\Delta} \zeta_p(3\Delta-n) \right]
\end{align*}
%\end{equation*}
%%
\item 
$m_1=0$ and $m_2=0$, means that $b_1$ coincides $c_1$ and $b_2$ with $c_2$.
\begin{equation*}
u^\Delta\left[1+ \left(p^n-2\right)\left(\sum_{l=1}^\infty p^{n(l-1)} p^{-3\Delta l}\right)\right]^2
= u^{\Delta}\,\left[ (1-2p^{-3\Delta})\zeta_p(3\Delta - n)\right]^2
\end{equation*}
\item 
$m_1, m_2\ne 0$ but $m_1+m_2 = d(c_1,c_2)$. This means that $b_1$ coincides with $b_2$, 
but neither of these coincides with either $c_1$ or $c_2$. 
\begin{align*}
& u^{2\Delta}(d-1)\,\left[ 2\left(p^n-1\right) \left(\sum_{l=1}^\infty p^{n(l-1)} p^{-3\Delta l}\right)
+ 2\left({p^n-1\atop 2}\right) \left(\sum_{l=1}^\infty p^{n(l-1)} p^{-3\Delta l}\right)^2\right]\\
&=\, - u^{2\Delta}\left(\log_p u + 1\right)\, (p^n-1)p^{-3\Delta}\zeta_p(3\Delta - n)\,\left[2 + 
(p^n-2) p^{-3\Delta} \zeta_p(3\Delta-n) \right]
\end{align*}
%%.
\item 
$m_1=0$ but $m_2\ne 0, d(c_1,c_2)$, which means $b_1$ coincides with $c_1$ but $b_2$ varies
(or $b_2$ coincides with $c_2$ but $b_1$ varies). 
\begin{align*}
& 2u^\Delta\left(\sum_{m_2=1}^{d(c_1,c_2)-1} p^{-m_2\Delta}\right) \left(1+ \sum_{l_1=1}^\infty(p^n-1) 
p^{n(l_1 -1)} p^{-3\Delta l_1}\right) \left(1+ \sum_{l_2=1}^\infty(p^n - 2) p^{n(l_2-1)} p^{-3\Delta l_2}\right)\\
& =\, 2 \left(p^{-\Delta}u^{\Delta} - u^{2\Delta}\right)\,\left(1-2p^{-3\Delta}\right)\zeta_p^2(3\Delta-n)
\frac{\zeta_p(\Delta)}{\zeta_p(3\Delta)}
\end{align*}
\item
$m_1,m_2\ne 0$ and $2\le m_1+m_2 < d(c_1,c_2)$. This means that both $b_1$ and $b_2$ are strictly 
inside the segment  $\langle c_1,c_2\rangle$, but $b_1$ is always to the left of $b_2$. 
\begin{align*} 
& u^\Delta\left(\sum_{m_2=1}^{d(c_1,c_2)-2}p^{-m_2\Delta}\left(\sum_{m_1=1}^{d(c_1,c_2)-m_2-1}
p^{-m_1\Delta}\right)\right)\, \left[1+\sum_{l=1}^\infty (p^n-1)p^{n(l-1)} p^{-3\Delta l}\right]^2 \\
& =\,\Big( u^{2\Delta}\left(\log_p u + 2 - \zeta_p(\Delta)\right) + p^{-2\Delta}\zeta_p(\Delta)  u^\Delta 
\Big) \left(\frac{\zeta_p(3\Delta - n)}{\zeta_p(3\Delta)}\right)^2\zeta_p(\Delta)
\end{align*}
\end{itemize}
The sums being geometric series could be performed easily.
 
There are additional diagrams with the positions of $a_1,b_1$ and $a_2,b_2$ interchanged,
however, some of the degenerate cases of these are indistinguishable. Exchanging the labels 
1 and 2 in W$_1$ in \cref{subway1-4pf}, we get $\widetilde{\mathrm{W}}_1$, which
contributes
\begin{align}
\hat{\mathcal{A}}_4^{(3)}(\widetilde{\mathrm{W}}_1) 
&=  \Big\{
u^{5\Delta}\, (1-2p^{-3\Delta})^2 +  2 \left(p^{-3\Delta}u^{2\Delta} - u^{5\Delta}\right)\,
\left(1-2p^{-3\Delta}\right)\\
&\qquad\qquad -\,  u^{2\Delta}\Big(\left(\log_p u + 2 \right) \frac{p^{-3\Delta}}{\zeta_p(3\Delta)} 
+ p^{-6\Delta}\Big) + u^{5\Delta}\Big\}\,\zeta^2_p(3\Delta - n)\\
& = \left\{ - u^{2\Delta}\Big(\left(\log_p u + 4\right)\frac{p^{-3\Delta}}{\zeta_p(3\Delta)} - p^{-6\Delta}\Big)
+ 4 p^{-6\Delta} u^{5\Delta}\right\}\,\zeta^2_p(3\Delta - n) 
\label{b4_B3_tW2}
\end{align}
where we have taken care to avoid double counting.

We draw attention to the fact that the functional dependence on the (single) cross 
ratio $u$ of  the CFT four-point correlation function, with contributions from the bulk 
3-vertex, involves terms with $u^{2\Delta}$ and $u^{2\Delta}\log_{p} u$, familiar from
the contribution from the contact interaction. However, there are additional powers 
$u^\Delta$ and $u^{5\Delta}$, although these do not accompany terms with 
$\log_p u$. We shall come back to this.

%%%%%%%%%%%%%%%%

%%%%%%%%%%%%%%%%%%

\begin{figure}[h]
\centering
\subfigure{
\begin{tikzpicture}[scale=0.7,every node/.style={scale=0.7}]
%\draw[step=1cm,gray,thin] (-3,-3) grid (3,3);
\draw (0,0) circle (3cm) ;
\filldraw[thick] (-1,0) -- (1,0);
\filldraw[thick] (-1,0) -- (-2.25,2);
\filldraw[thick] (-1,0) -- (-2.25,-2);
\filldraw[thick] (1,0) -- (2.25,2);
\filldraw[thick] (1,0) -- (2.25,-2);
\filldraw[thick] (-1.6,1) -- (-.4,1.7);
\filldraw[thick] (1.6,1) -- (.4,1.7);
%\filldraw[thick] (-.6,-1.5) -- (0,-2.4);
\node[mark size=2pt,color=red] at (-1,0) {\pgfuseplotmark{*}};
\node[mark size=2pt,color=blue] at (-1.6,1) {\pgfuseplotmark{*}};
\node[mark size=2pt,color=blue] at (-.4,1.7) {\pgfuseplotmark{*}};
\node[mark size=2pt,color=blue] at (1.6,1) {\pgfuseplotmark{*}};
\node[mark size=2pt,color=blue] at (.4,1.7) {\pgfuseplotmark{*}};
\node[mark size=2pt,color=red] at (1,0) {\pgfuseplotmark{*}};
\node at (-2.33,2.25) {$x_{1} $};
\node at (-2.33,-2.25) {$x_{2} $};
\node at (2.33,2.25) {$x_{3} $};
\node at (2.33,-2.25) {$x_{4} $};
\node at (3.0,-2.5) {W$_4$};
\end{tikzpicture}
}
\hspace{36pt}
\subfigure{
\begin{tikzpicture}[scale=0.7,every node/.style={scale=0.7}]
%\draw[step=1cm,gray,thin] (-3,-3) grid (3,3);
\draw (0,0) circle (3cm) ;
\filldraw[thick] (-1,0) -- (1,0);
\filldraw[thick] (-1,0) -- (-2.25,2);
\filldraw[thick] (-1,0) -- (-2.25,-2);
\filldraw[thick] (1,0) -- (2.25,2);
\filldraw[thick] (1,0) -- (2.25,-2);
\filldraw[thick] (-1.6,1) -- (-.4,1.7);
\filldraw[thick] (-1.6,-1) -- (-.4,-1.7);
%\filldraw[thick] (-.6,-1.5) -- (0,-2.4);
\node[mark size=2pt,color=red] at (-1,0) {\pgfuseplotmark{*}};
\node[mark size=2pt,color=blue] at (-1.6,1) {\pgfuseplotmark{*}};
\node[mark size=2pt,color=blue] at (-.4,1.7) {\pgfuseplotmark{*}};
\node[mark size=2pt,color=blue] at (-1.6,-1) {\pgfuseplotmark{*}};
\node[mark size=2pt,color=blue] at (-.4,-1.7) {\pgfuseplotmark{*}};
\node[mark size=2pt,color=red] at (1,0) {\pgfuseplotmark{*}};
\node at (-2.33,2.25) {$x_{1} $};
\node at (-2.33,-2.25) {$x_{2} $};
\node at (2.33,2.25) {$x_{3} $};
\node at (2.33,-2.25) {$x_{4} $};
\node at (3.5,-2.5) {W$_5$};
\end{tikzpicture}
}\\
\subfigure{
\begin{tikzpicture}[scale=0.7,every node/.style={scale=0.7}]
%\draw[step=1cm,gray,thin] (-3,-3) grid (4,3);
\draw (0,0) circle (3cm) ;
\filldraw[thick] (-1,0) -- (1,0);
\filldraw[thick] (-1,0) -- (-2.25,2);
\filldraw[thick] (-1,0) -- (-2.25,-2);
\filldraw[thick] (1,0) -- (2.25,2);
\filldraw[thick] (1,0) -- (2.25,-2);
\filldraw[thick] (-1.2,-.3) -- (-2,.5);
\filldraw[thick] (-1.6,-.95) -- (-2.55,-.1);
\node[mark size=2pt,color=red] at (-1,0) {\pgfuseplotmark{*}};
\node[mark size=2pt,color=blue] at (-1.2,-.3) {\pgfuseplotmark{*}};
\node[mark size=2pt,color=blue] at (-1.6,-.95) {\pgfuseplotmark{*}};
\node[mark size=2pt,color=blue] at (-2,.5) {\pgfuseplotmark{*}};
\node[mark size=2pt,color=blue] at (-2.55,-.1) {\pgfuseplotmark{*}};
\node[mark size=2pt,color=red] at (1,0) {\pgfuseplotmark{*}};
\node at (-2.33,2.25) {$x_{1} $};
\node at (-2.33,-2.25) {$x_{2} $};
\node at (2.33,2.25) {$x_{3} $};
\node at (2.33,-2.25) {$x_{4} $};
\node at (3.5,-2.5) {W$_6$};
\end{tikzpicture}
}
\hspace{36pt}
\subfigure{
\begin{tikzpicture}[scale=0.7,every node/.style={scale=0.7}]
%\draw[step=1cm,gray,thin] (-3,-3) grid (3,3);
\draw (0,0) circle (3cm) ;
\filldraw[thick] (-1,0) -- (1,0);
\filldraw[thick] (-1,0) -- (-2.25,2);
\filldraw[thick] (-1,0) -- (-2.25,-2);
\filldraw[thick] (1,0) -- (2.25,2);
\filldraw[thick] (1,0) -- (2.25,-2);
\filldraw[thick] (-1.6,-.95) -- (-.6,-1.5);
\filldraw[thick] (-.6,-1.5) -- (.4,-1.5);
\filldraw[thick] (-.6,-1.5) -- (0,-2.4);
\node[mark size=2pt,color=red] at (-1,0) {\pgfuseplotmark{*}};
\node[mark size=2pt,color=blue] at (.4,-1.5) {\pgfuseplotmark{*}};
\node[mark size=2pt,color=blue] at (-1.6,-.95) {\pgfuseplotmark{*}};
\node[mark size=2pt,color=blue] at (0,-2.4) {\pgfuseplotmark{*}};
\node[mark size=2pt,color=blue] at (-.6,-1.5) {\pgfuseplotmark{*}};
\node[mark size=2pt,color=red] at (1,0) {\pgfuseplotmark{*}};
\node at (-2.33,2.25) {$x_{1} $};
\node at (-2.33,-2.25) {$x_{2} $};
\node at (2.33,2.25) {$x_{3} $};
\node at (2.33,-2.25) {$x_{4} $};
\node at (3.5,-2.5) {W$_7$};
\end{tikzpicture}
}
\caption{\small
Other Witten (subway) diagrams with bulk 3-point interactions that give
trivial dependence on the cross ratio.}
\label{subway3-4pf}
\end{figure}
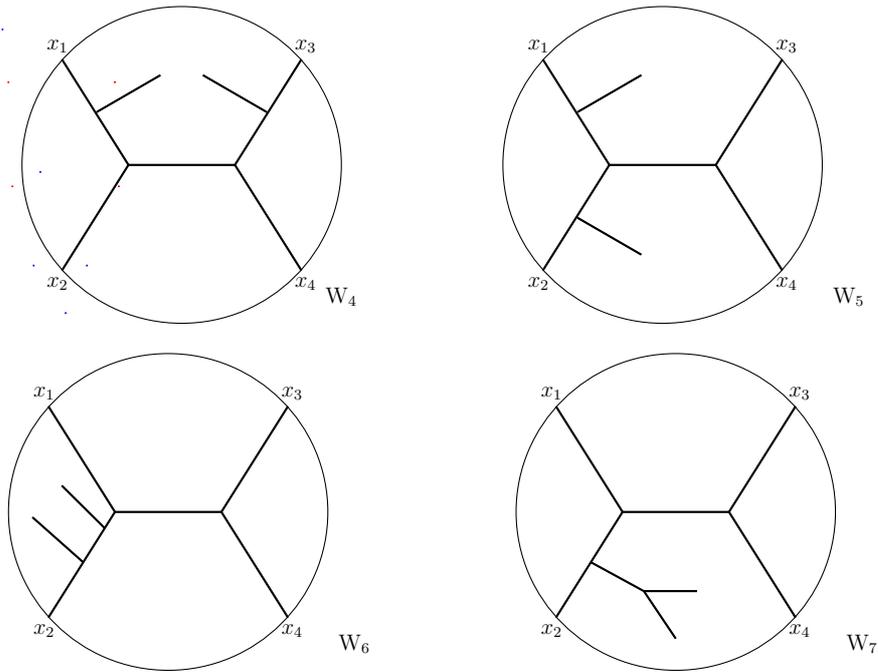

Many more Witten diagrams contribute to the boundary four-point amplitude. These are
shown in \cref{subway2-4pf,subway3-4pf} without labels, for which there are more than
one possibilities for each diagram. A `twig' can be attached to an external in many 
different ways, however, we have not shown all these diagrams explicitly. In case an
exchange of the vertices 1 and 2 lead to different configurations, we denote it with a tilde.

When a twig is attached to an external leg, say, as in the diagram W$_3$ in the 
configuration in which the twig $\left\langle b_1,a_1\right\rangle$ is attached to the 
external leg to $x_1$, it involves $K(x_1,a_1) = K(x_1,b_1) G(b_1,a_1)$. In order to 
perform the sum over the interaction point, it is simpler to use the fact that the 
bulk-to-boundary propagator is a (regularised) limit of a bulk-to-bulk propagator: 
$K(x_1,b_1) = \displaystyle{\lim_{z_1\to x_1}} G(z_1,b_1)$. The same limiting procedure
is used in calculating the contributions from $\widetilde{\mathrm{W}}_3$ and those in  
\cref{subway3-4pf}.

We shall skip the details of each individual cases and present only the results. The 
relevant expressions corresponding to \cref{subway2-4pf,subway3-4pf} are
%%%
\begin{align}
\hat{\mathcal{A}}_{4}^{(3)}\left(\mathrm{W}_2\right) &= - {p^{-7\Delta+n}} 
\Big( (p^n-1)\log_p u  - (p^n-3) \Big) \left(2 + (p^n-1) p^{-3\Delta} 
\zeta_p(3\Delta-n) \right)\\
&\qquad\qquad\qquad\qquad\times\, 
{\zeta_p(3\Delta - n)\zeta_p(4\Delta - n)}\, u^{2\Delta}\\
\hat{\mathcal{A}}_{4}^{(3)}\big(\mathrm{W}_3\big) &=
 4 p^{-\Delta}  \frac{\zeta_p(\Delta)\zeta_p^2(3\Delta-n)}{\zeta_p(3\Delta)}
 \left\{ \frac{\zeta_p(\Delta)}{\zeta_p(3\Delta)}\Big( p^{-\Delta} u^{\Delta}
 - u^{2\Delta}\Big) + \left(1-2p^{-3\Delta}\right)\Big(u^\Delta + u^{2\Delta}\Big)
 \right\}\\
\hat{\mathcal{A}}_{4}^{(3)}\big(\widetilde{\mathrm{W}}_3\big) &=
 4 p^{-3\Delta}   \frac{\zeta_p^2(3\Delta-n)}{\zeta_p(3\Delta)}
 \Big\{ u^{2\Delta} -  2p^{-3\Delta}\zeta_p(3\Delta) u^{5\Delta} \Big\} \\
\hat{\mathcal{A}}_{4}^{(3)}\big(\mathrm{W}_4 + \widetilde{\mathrm{W}}_4\big) &=
 4 p^{-2\Delta}  \frac{\zeta_p^2(3\Delta-n)}{\zeta^2_p(3\Delta)}
 \Big\{ \zeta_p^2(\Delta) u^{\Delta} +  p^{-4\Delta} \zeta^2_p(3\Delta) u^{5\Delta}\Big\}  \\
\hat{\mathcal{A}}_{4}^{(3)}\left(\mathrm{W}_5\right) &=
 4 p^{-4\Delta}  \frac{\zeta_p(\Delta)\zeta_p^2(3\Delta-n)}{\zeta_p(3\Delta)}\,  u^{2\Delta}\\
\hat{\mathcal{A}}_{4}^{(3)}\big(\mathrm{W}_6 + \widetilde{\mathrm{W}}_6\big) &= 
 4 p^{-3\Delta}  \frac{\zeta_p(2\Delta)\zeta_p^2(3\Delta-n)}{\zeta_p^2(3\Delta)}
 \bigg(\zeta_p(\Delta) + p^{-2\Delta}\zeta_p(3\Delta) \bigg)\,  u^{2\Delta}\\
\hat{\mathcal{A}}_{4}^{(3)}\left(\mathrm{W}_7\right) &=
  4p^{-5\Delta}(p^n-1)\zeta_p(2\Delta)\zeta_p(3\Delta-n)
 \Bigg\{ \left(2 + (p^n-2) p^{-3\Delta} \zeta_p(3\Delta-n) \right) \\
 &\qquad\qquad\qquad + p^{-4\Delta + n}\zeta_p(4\Delta-n)
 \left(2 + (p^n-1) p^{-3\Delta} \zeta_p(3\Delta-n) \right)\Bigg\}\, u^{2\Delta}.
\label{b4_B3_rest}
\end{align}
%%%%%
The boundary four-point correlator, computed from the bulk theory with three-scalar interaction 
vertex, is obtained by adding all the contributions discussed above. Combined with the result 
\cref{contact4} from the bulk four-point contact term in \cite{Gubser:2016guj}, these exhaust the 
possible bulk contributions to the boundary four-point amplitude. It may be possible to simplify 
the result further, but it is unlikely to be very useful. 

We see that some of the diagrams contribute terms with $u^{5\Delta}$, which would signify the 
presence of an operator of dimensions $5\Delta$ in the intermediate channel. However, this is 
not expected. Indeed, when we collect the terms in \cref{b4_B3_tW2,b4_B3_rest},
%%
%\begin{equation*}
%\zeta_p^2(3\Delta-n) \Big[ \left(1-2p^{-3\Delta}\right)^2 - 2\left(1-2p^{-3\Delta}\right) + 1
%- 8p^{-6\Delta}  + 4p^{-6\Delta}\Big] = 0 
%\end{equation*}
the coefficient of $u^{5\Delta}$ vanishes.

%%%%%%%%%%%%%%%%%%
%%%%%%%%%%%%%%%%%%%%%%%%%%%%%%%%%

%%%%%%%%%%%%%%%%%%%%%%%%%%%%%%%%%%%%%
\section{Correlators in Mellin space}\label{sec:mellin}
Mellin space is the natural setting for conformally invariant field theories 
\cite{Liu:1998th,Mack:2009gy} (see \cite{Penedones:2016voo} for a review). The isomorphism 
between the additive group $(\mathbb{R},+)$ and the multiplicative group $(\mathbb{R}^+,\times)$,
via the exponential map, is at the heart of this idea --- radial quantization maps the translation 
invariance of a field theory to scale invariance of CFT, and takes Fourier transform to Mellin 
transform. The generic branch point singularities of a CFT correlator takes the form of (isolated) 
poles in terms of the Mellin variables. Many authors have explored CFTs from a Mellin space 
perspective in recent times. For a partial list of references dealing with different aspects, see 
\cite{Penedones:2010ue,Fitzpatrick:2011ia,Paulos:2011ie,Costa:2012cb,Costa:2014kfa,
Rastelli:2016nze,Gopakumar:2016wkt}.

Let us consider the results obtained in \cref{sec:A4_3vertex}. In the $s$-channel, the cross 
ratio defined in \cref{xratio}, in the four point correlation function of the boundary CFT, evaluated 
from the bulk with both the three- and four-point interaction terms, is $u<1$. The amplitude has 
the following form
\begin{align}
\mathcal{A}_4(s) &= \left(\prod_{i=1,2} K(x_{i},c_{1})\prod_{j=3,4} K(x_{j},c_{2}) \right)
\widetilde{\mathcal{A}}_4(s)\nonumber\\
\widetilde{\mathcal{A}}_4(s) &= \alpha_{(1)} u^\Delta + \left(\alpha_{(2)} + \beta_{(2)}\log_p u\right)
u^{2\Delta},
\label{amps4Schematic}
\end{align}
where the coefficients $\alpha_{(1)}$, $\alpha_{(2)}$ and $\beta_{(2)}$, which depend only on 
$p$, $\Delta$, $g_3$ and $g_4$, can be determined from the. results in 
\cref{sec:UMHol,sec:A4_3vertex}.

In the $s$-channel discussed above, we have considered the topology (shown in \cref{subway1-4pf}) 
in which path from $x_{1}$ to $x_{2}$ meet for the first time at $c_{1}$, and that from $x_{3}$ to 
$x_{4}$ at $c_2$.  There is another possibility in which the path from $x_{1}$ to $x_{3}$ and $x_{2}$ 
to $x_{4}$ meet for the first time. The amplitude for this case, i.e., for the $t$-channel, can be obtained 
by a simple relabelling of the external points. Therefore, it depends on the cross ratio 
\begin{equation*}
v=\bigg|\frac{x_{13}x_{24}}{x_{12}x_{34}}\bigg|_{q}=\frac{1}{u}
\end{equation*}
Notice that $v$ is not an independent variable, since there is only one independent cross ratio, as
emphasised in the \cref{sec:UMHol}. Therefore, the $t$-channel amplitude is
%%
%\begin{equation}
\begin{align}
\mathcal{A}_4(t) &= \left(\prod_{i=1,3} K(x_{i},c_{1})\prod_{j=2,4} K(x_{j},c_{2}) \right)
\widetilde{\mathcal{A}}_4(t)\\
\widetilde{\mathcal{A}}_4(t) &= \alpha_{(1)} u^{-\Delta} + \left(\alpha_{(2)} - \beta_{(2)}\log_p u\right)
u^{-2\Delta} ,
\label{ampt4Schematic}
\end{align} 
%\end{equation}
%%
which has the same form as \cref{amps4Schematic} with some flips in sign.

Let us define the Mellin amplitude\footnote{There is a notion of $p$-adic Mellin transformation 
that uses the map from $\Qp$ to $\Qp^*$. However, the scalar field is complex valued and since
we are considering complex valued functions on ultrametric spaces, the Mellin transform here is 
the standard one.}
\begin{equation}
\widetilde{\mathcal{A}}_4\left(q;u\right) = \frac{1}{2\pi i} \int_{-i\infty}^{i\infty} u^{z}\mathcal{M}_4(z)\,dz,
\label{def-mellin}
\end{equation}
where the amplitude in the $q$-channel ($q=s,t$) can be obtained by closing the contour on the right- 
or the left-half plane depending on whether $u<1$ or $u>1$. We have not included any conventional 
factor of $\Gamma$-functions in the definition above because in this situation, where there is only
one independent cross ratio, and the dynamics is rather simple, this does not seem necessary.

The generic form of the Mellin amplitude is then
\begin{equation}
\mathcal{M}_4(z) = \frac{a_{(1)}}{z^{2} - \Delta^{2}} + \frac{a_{(2)}}{z^{2} -  4\Delta^{2}} + 
\frac{b_{(2)}}{\left(z^{2} - 4\Delta^{2}\right)^{2}}.    
\label{mellin-poles}
\end{equation}
It can be seen that this leads to the terms in \cref{amps4Schematic,ampt4Schematic} with the 
correct relative signs of the different terms. The actual coefficients can be determined by a detailed 
comparison. We see that the operator of dimension $\Delta$ dual to the bulk scalar field, as well 
as the composite operator of dimension $2\Delta$, mediate the CFT 4-point amplitude. We also 
find that the $\log_p u$ term only comes with $u^{2\Delta}$. (We find this to be true also of 
the terms in the five-point amplitudes that we have evaluated.) This is what one expects to 
find in this theory. On the external legs, we have the primary operator of dimension $\Delta$ 
of the boundary CFT. Therefore, the condition \cite{Liu:1998th,Gopakumar:2016wkt} for the 
appearance of a logarithm is met only when the dimension of the mediating operator is 
$2\Delta$.

In summary, $p$-adic CFT holographically dual to a bulk scalar field theory, takes a rather 
simple form in Mellin space, reflecting the absence of the tower of secondary operators.

%%%%%%%%%%%%%%%%%%
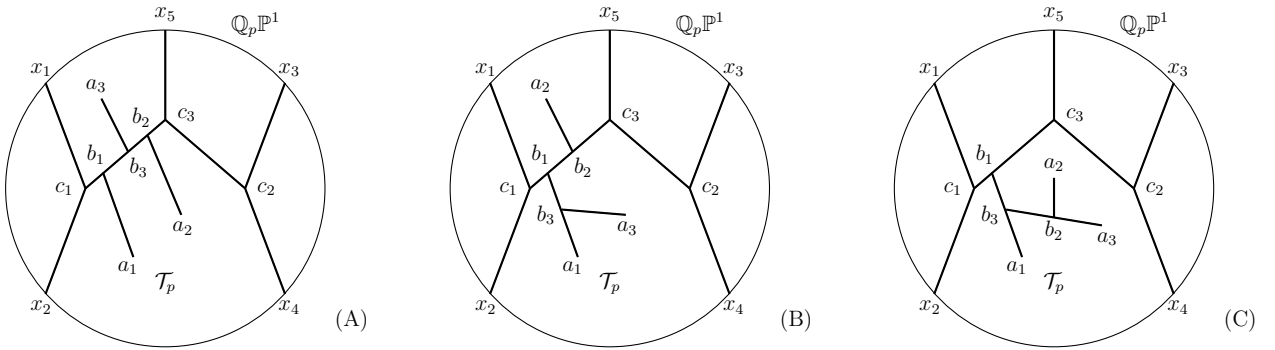
\begin{figure}[H]
\centering
\subfigure{
\begin{tikzpicture}[scale=0.7,every node/.style={scale=0.7}]
%\draw[step=1cm,gray,thin] (-3,-3) grid (3,3);
\draw (0,0) circle (3cm) ;
\filldraw[thick] (0,3) -- (0,1.3);
\filldraw[thick] (-1.5,0) -- (-2.25,2);
\filldraw[thick] (-1.5,0) -- (-2.25,-2);
\filldraw[thick] (1.5,0) -- (2.25,2);
\filldraw[thick] (1.5,0) -- (2.25,-2);

\filldraw[thick] (-1.5,0) -- (0,1.3);
\filldraw[thick] (1.5,0) -- (0,1.3);

\filldraw[thick] (-1.17,0.3) -- (-0.6,-1.3);
\filldraw[thick] (-.7,.7) -- (-1.2,1.7);
\filldraw[thick] (-0.33,1) -- (0.3,-0.5);
%\filldraw[thick] (-1,0) -- (-1,-1.3);
\node[mark size=2pt,color=red] at (-1.5,0) {\pgfuseplotmark{*}};
\node[mark size=2pt,color=blue] at (-1.17,0.3) {\pgfuseplotmark{*}};
\node[mark size=2pt,color=blue] at (-.6,-1.3) {\pgfuseplotmark{*}};
\node[mark size=2pt,color=blue] at (-0.7,0.7) {\pgfuseplotmark{*}};
\node[mark size=2pt,color=blue] at (-1.2,1.7) {\pgfuseplotmark{*}};
\node[mark size=2pt,color=blue] at (-0.33,1) {\pgfuseplotmark{*}};
\node[mark size=2pt,color=blue] at (0.3,-0.5) {\pgfuseplotmark{*}};
\node[mark size=2pt,color=red] at (1.5,0) {\pgfuseplotmark{*}};
\node[mark size=2pt,color=red] at (0,1.3) {\pgfuseplotmark{*}};
\node at (-1.9,0) {$c_{1}$};
\node at (0.4,1.4) {$c_{3}$};
\node at (1.9,0) {$c_{2}$};

\node at (-.5,0.4) {$b_{3}$};
\node at (-1.3,0.65) {$b_{1}$};
\node at (-0.45,1.35) {$b_{2}$};

\node at (-1.3,1.95) {$a_{3}$};
\node at (-0.7,-1.5) {$a_{1}$};
\node at (0.33,-0.75) {$a_{2}$};

\node at (-2.33,2.25) {$x_{1} $};
\node at (-2.33,-2.25) {$x_{2} $};
\node at (2.33,2.25) {$x_{3} $};
\node at (2.33,-2.25) {$x_{4} $};
\node at (0,3.3) {$x_{5} $};

\node at (0.0,-1.8) {$\mathcal{T}_{p}$};
\node at (1.7,3.1) {$\mathbb{Q}_{p}\mathbb{P}^{1}$};
\node at (3.5,-2.5) {(A)};
\end{tikzpicture}
}
\hspace{12pt}
\subfigure{
\begin{tikzpicture}[scale=0.7,every node/.style={scale=0.7}]
%\draw[step=1cm,gray,thin] (-3,-3) grid (3,3);
\draw (0,0) circle (3cm) ;
\filldraw[thick] (0,3) -- (0,1.3);
\filldraw[thick] (-1.5,0) -- (-2.25,2);
\filldraw[thick] (-1.5,0) -- (-2.25,-2);
\filldraw[thick] (1.5,0) -- (2.25,2);
\filldraw[thick] (1.5,0) -- (2.25,-2);

\filldraw[thick] (-1.5,0) -- (0,1.3);
\filldraw[thick] (1.5,0) -- (0,1.3);

\filldraw[thick] (-1.17,0.3) -- (-0.6,-1.3);
\node[mark size=2pt,color=blue] at (-1.17,0.3) {\pgfuseplotmark{*}};
\node[mark size=2pt,color=blue] at (-.6,-1.3) {\pgfuseplotmark{*}};
\filldraw[thick] (-.7,.7) -- (-1.2,1.7);
\node[mark size=2pt,color=blue] at (-0.7,0.7) {\pgfuseplotmark{*}};
\node[mark size=2pt,color=blue] at (-1.2,1.7) {\pgfuseplotmark{*}};
\filldraw[thick] (-0.9,-0.4) -- (0.3,-0.5);
\node[mark size=2pt,color=blue] at (-0.9,-0.4) {\pgfuseplotmark{*}};
\node[mark size=2pt,color=blue] at (0.3,-0.5) {\pgfuseplotmark{*}};

\node at (-.5,0.4) {$b_{2}$};
\node at (-1.3,0.65) {$b_{1}$};
\node at (-1.2,-0.5) {$b_{3}$};

\node at (-1.3,1.95) {$a_{2}$};
\node at (-0.7,-1.5) {$a_{1}$};
\node at (0.33,-0.75) {$a_{3}$};

\node[mark size=2pt,color=red] at (-1.5,0) {\pgfuseplotmark{*}};
\node[mark size=2pt,color=red] at (1.5,0) {\pgfuseplotmark{*}};
\node[mark size=2pt,color=red] at (0,1.3) {\pgfuseplotmark{*}};
\node at (-1.9,0) {$c_{1}$};
\node at (0.4,1.4) {$c_{3}$};
\node at (1.9,0) {$c_{2}$};

\node at (-2.33,2.25) {$x_{1} $};
\node at (-2.33,-2.25) {$x_{2} $};
\node at (2.33,2.25) {$x_{3} $};
\node at (2.33,-2.25) {$x_{4} $};
\node at (0,3.3) {$x_{5} $};

\node at (0.0,-1.8) {$\mathcal{T}_{p}$};
\node at (1.7,3.1) {$\mathbb{Q}_{p}\mathbb{P}^{1}$};
\node at (3.5,-2.5) {(B)};
\end{tikzpicture}
}
\hspace{12pt}
\subfigure{
\begin{tikzpicture}[scale=0.7,every node/.style={scale=0.7}]
%\draw[step=1cm,gray,thin] (-3,-3) grid (3,3);
\draw (0,0) circle (3cm) ;
\filldraw[thick] (0,3) -- (0,1.3);
\filldraw[thick] (-1.5,0) -- (-2.25,2);
\filldraw[thick] (-1.5,0) -- (-2.25,-2);
\filldraw[thick] (1.5,0) -- (2.25,2);
\filldraw[thick] (1.5,0) -- (2.25,-2);

\filldraw[thick] (-1.5,0) -- (0,1.3);
\filldraw[thick] (1.5,0) -- (0,1.3);

\filldraw[thick] (-1.17,0.3) -- (-0.6,-1.3);
\node[mark size=2pt,color=blue] at (-1.17,0.3) {\pgfuseplotmark{*}};
\node[mark size=2pt,color=blue] at (-.6,-1.3) {\pgfuseplotmark{*}};
\filldraw[thick] (0,-.55) -- (0,0.2);
\node[mark size=2pt,color=blue] at (-0,-0.55) {\pgfuseplotmark{*}};
\node[mark size=2pt,color=blue] at (0,0.2) {\pgfuseplotmark{*}};
\filldraw[thick] (-0.9,-0.4) -- (0.9,-0.7);
\node[mark size=2pt,color=blue] at (-0.9,-0.4) {\pgfuseplotmark{*}};
\node[mark size=2pt,color=blue] at (0.9,-0.7) {\pgfuseplotmark{*}};

\node at (0,-0.79) {$b_{2}$};
\node at (-1.3,0.65) {$b_{1}$};
\node at (-1.2,-0.5) {$b_{3}$};

\node at (0,0.45) {$a_{2}$};
\node at (-0.7,-1.5) {$a_{1}$};
\node at (1.0,-0.95) {$a_{3}$};

\node[mark size=2pt,color=red] at (-1.5,0) {\pgfuseplotmark{*}};
\node[mark size=2pt,color=red] at (1.5,0) {\pgfuseplotmark{*}};
\node[mark size=2pt,color=red] at (0,1.3) {\pgfuseplotmark{*}};
\node at (-1.9,0) {$c_{1}$};
\node at (0.4,1.4) {$c_{3}$};
\node at (1.9,0) {$c_{2}$};

\node at (-2.33,2.25) {$x_{1} $};
\node at (-2.33,-2.25) {$x_{2} $};
\node at (2.33,2.25) {$x_{3} $};
\node at (2.33,-2.25) {$x_{4} $};
\node at (0,3.3) {$x_{5} $};

\node at (0.0,-1.8) {$\mathcal{T}_{p}$};
\node at (1.7,3.1) {$\mathbb{Q}_{p}\mathbb{P}^{1}$};
\node at (3.5,-2.5) {(C)};
\end{tikzpicture}
}
\caption{A few Witten diagrams contributing to five-point amplitude in CFT.}
\label{fivepoint}
\end{figure}
%%%%%%%%%%%%%%%%%%%%%%%%%%%%%
 
%%%%%%%%%%%%%%%%%%%%%%%%%%%%%%%%%%%%%%%%
\section{Conclusions}\label{sec:concl}
We have analyzed some aspects of CFTs on ultrametric spaces in the recently proposed 
holographic approach. The natural relation between the $p$-adic field (or its algebraic 
extensions) as the boundary of an infinite lattice, the so called Bruhat-Tits tree, as the 
bulk, is essential in this correspondence. In the simplest case, the bulk theory is described
by the naturally discretized action of an interacting scalar field. Two-, three- and four-point
correlators of the boundary CFT were already computed with bulk-to-boundary and 
bulk-to-bulk propagators, and the structural similarity of these propagators, as well as 
the resulting amplitudes, with the usual Archimedean CFTs were emphasised in 
Refs.\cite{Gubser:2016guj,Heydeman:2016ldy}. 

However, in computing the four-point correlator, contributions from only the bulk four-scalar 
interaction was used. We have computed the exchange contributions from the bulk three-scalar 
interaction. Since there cannot be any other contribution to the CFT four-point amplitude, this
gives the complete answer. Our results show that the fundamental operator of dimension
$\Delta$ and a composite of dimension $2\Delta$ flow in the intermediate channels. The
latter has an anomalous contribution to its dimension resulting in a term dependent on the
logarithm of the cross ratio. 

The singularities, when viewed in the Mellin variables conjugate to the cross ratios, result 
in simple poles at $\pm\Delta$ and $\pm 2\Delta$. Moreover, there are double poles at 
$\pm2\Delta$ which accounts for the term with $u^{2\Delta}\log_p u$. 

It is possible to compute higher-point amplitudes in CFT, although it gets rapidly complicated
due to contributions from a large number of possible diagrams. For instance, a few of the 
many possible configurations of the boundary five-point correlator is shown in \cref{fivepoint}. 
We have chosen the boundary points $x_1,\cdots,x_5$ in such a way that the  propagators 
from $x_{1}$ and $x_{2}$ to the interaction vertex at $a_1$ meet first at $c_{1}$. Similarly 
those joining $x_{3}$ and $x_{4}$ to $a_2$ meet at $c_{2}$. Finally, $x_{5}$ and the fields 
in the intermediate channel meet at the vertex $a_3$. In this process the path from $x_5$
meets that joining $c_1$ and $c_2$ at $c_{3}$. The five-point amplitude is a function 
of two cross ratios, which are given by the lengths of the paths $\left\langle c_{1},c_{3} 
\right\rangle$ and $\left\langle c_{2},c_{3} \right\rangle$ . 
The decomposition of the propagators and the counting proceeds in the same way as 
for the four-point correlator explained in \cref{sec:A4_3vertex}. As an example, let us 
consider the diagram in the left-most panel in \cref{fivepoint}, and take the $b$-nodes 
strictly between the vertices $c_1$ and $c_3$. In this configuration, the contribution of 
the Witten diagram can be decomposed as follows
\begin{align}
\label{amp:5pt}
&g_3^3\,\Big( K(x_1,c_1) K(x_2,c_1) K(x_3,c_2)K(x_4,c_2)
K(x_5,c_3)\Big) \,G^2(c_2,c_3)\\
&\quad \times \sum_{a_1,a_2,a_3} \underbrace{\Big(G^2(c_{1},b_{1})
G(b_{1},b_{3})G^2(b_{3},b_{2})G^3(b_2,c_3)\Big)}_{\textbf{I}}\,
\underbrace{\Big( G(b_{1},a_{1})G(b_{2},a_{2})G(b_{3},a_{3})\Big)^3}_{\textbf{II}}.
\end{align}
The contribution from the sum of propagators in \textbf{II} does not depend on the 
cross ratios yielding factors depending only on $p$ and $\Delta$. This factor, for this
diagram, is $\left(\displaystyle{\frac{\zeta_p(3\Delta-n)}{\zeta_p(3\Delta)}}\right)^3$. 
The only non-trivial dependence on the cross ratios comes from \textbf{I} and its
equivalent for the three diagrams in \cref{fivepoint}, which are, respectively,
\begin{align}
\textbf{I}(\textrm{A}) &=\: - p^{-4\Delta}\zeta_p^2(\Delta)\Big(\zeta_p(\Delta) - p^{-\Delta}
\zeta_p(2\Delta)\Big) u_1^{\Delta} - \zeta_p^2(\Delta)\zeta_p(2\Delta) u_1^{3\Delta}\\
&\:\qquad - \zeta^2_p(\Delta)\left(\log_p u_1 + p^{-\Delta}\left(p^{-\Delta} 
\frac{\zeta_p(2\Delta)}{\zeta_p(\Delta)} - \zeta_p(\Delta) + 3\right) \right) u_1^{2\Delta} \\
\textbf{I}(\textrm{B}) &=\:  -p^{-\Delta}\zeta_p(\Delta)\, \Big(\log_p u_1 + 2 + p^{-\Delta}\zeta_p(\Delta)
\Big) u_1^{2\Delta} + \zeta^2_p(\Delta) u_1^{3\Delta}\\
\textbf{I}(\textrm{C}) &=\:  p^{-\Delta}\zeta_p(\Delta)\, u_1^{3\Delta} - \zeta_p(\Delta)\, u_1^{4\Delta} ,  
\label{amp:5pt_2}
\end{align}
where $u_1 = p^{-d(c_1,c_3)}$ and $u_2 = p^{-d(c_2,c_3)}$ are the two cross ratios, (but 
the dependence on $G^{2}(c_{2},c_{3}) = u_2^{2\Delta}$ is not part of $\textbf{I}$). 
Operators of dimensions ${\Delta}$ and $2\Delta$, the latter accompanied also by terms 
with $\log_p u_1$, can be seen to contribute to the intermediate channel. These operators 
mediate interactions in CFT and give a closed operator algebra.

Finally, we have argued that there is little qualitative difference between CFTs on $p$-adic 
space or its finite algebraic extensions, even though the latter are finite dimensional vector 
spaces over $\Qp$. In either case, the bulk is a Bruhat-Tits tree on which paths joining nodes 
are unique. The restricted ultrametric geometry implies that the $n$-point correlator of the CFT 
is a function of $(n-3)$ cross ratios. In a sense, all these holographic CFTs are analogues 
of (Archimedean) adS$_2$/CFT$_1$. This is more apparent from its structure in the Mellin 
space. Be that as it may, the correspondence between the infinite tree and the ultrametric 
spaces on its boundary is inherently holographic. It is, therefore, a natural setting where 
one may address unresolved issues in holography and hope to get some interesting 
answers.

%%%%%%%%%%%%%%%%%%%%%%%%%%%%%%%%%
 
\bigskip\bigskip

\noindent{\bf Acknowledgments:} This work is supported by the research grant no.~5304-2, 
{\em Symmetries and Dynamics: Worldsheet and Spacetime}, from the Indo-French Centre 
for Promotion of Advanced Research (IFCPAR/CEFIPRA). We thank Sarthak Parikh for 
useful correspondence pointing out an error in the first version. 
 
%%%%%%%%%%%%%%%%%%%%%%%%%%%%%%%%%%%%

\bigskip

%\newpage

%%

%%%%%%%%%%%%%%%%%%%%%%%%%%%%%%%%%%%%%

\end{document}